\documentclass[prl,aps,twocolumn,superscriptaddress,showpacs,nobibnotes,epsf,graphicx]{revtex4}

\usepackage{graphicx}
\usepackage{dcolumn}
\usepackage{bm}
\usepackage{SIunits}
\usepackage{epstopdf}
\usepackage{array,color}
\usepackage{xspace}

\newcommand{\ie}{{\it i.e.}}
\newcommand{\Rb}{Rb$_2$Cr$_3$As$_3$\xspace}
\newcommand{\K}{K$_2$Cr$_3$As$_3$\xspace}
\newcommand{\A}{A$_2$Cr$_3$As$_3$\xspace}

\newcommand{\Tc}{$T_c$\xspace}
\newcommand{\Hc}{$H_{c2}$\xspace}
\newcommand{\Hperp}{$H_{\bot}$\xspace}
\newcommand{\Hpar}{$H_{\parallel}$\xspace}
\newcommand{\SP}{P$\overline{6}$m2\xspace}
\newcommand{\ClO}{(TMTSF)$_2$ClO$_4$\xspace}
\newcommand{\PF}{(TMTSF)$_2$PF$_6$\xspace}
\newcommand{\RNum}[1]{\uppercase\expandafter{\romannumeral #1\relax}}

\begin{document}

\title{Tunable electronic anisotropy in single-crystal A$_2$Cr$_3$As$_3$  \\
(A = K, Rb) quasi-one-dimensional superconductors}

\author{X. F.~Wang}
\author{C.~Roncaioli}
\author{C.~Eckberg}
\author{H.~Kim}
\author{J.~Yong}
\author{Y.~Nakajima}
\author{S.~R.~Saha}
\affiliation{Center for Nanophysics and Advanced Materials,
Department of Physics, University of Maryland, College Park, MD
20742}
\author{P.Y.~Zavalij}
\affiliation{Department of Chemistry, University of Maryland,
College Park, Maryland 20742, USA}
\author{J.~Paglione}
\affiliation{Center for Nanophysics and Advanced Materials,
Department of Physics, University of Maryland, College Park, MD
20742}
\affiliation{Canadian Institute for Advanced Research,
Toronto, Canada M5G 1Z8}

\date{\today}

\begin{abstract}

Single crystals of A$_2$Cr$_3$As$_3$ (A = K, Rb) were successfully grown using a self-flux method and studied via structural, transport and thermodynamic measurement techniques. The superconducting state properties between the two species are similar, with critical temperatures of 6.1 K and 4.8 K in \K and \Rb, respectively. However, the emergence of a strong normal state electronic anisotropy in \Rb suggests a unique electronic tuning parameter is coupled to the inter-chain spacing in the \A structure, which increases with alkali metal ionic size while the one-dimensional [(Cr$_{3}$As$_{3})^{2-}]_{\infty}$ chain structure itself remains essentially unchanged. Together with dramatic enhancements in both conductivity and magnetoresistance (MR), the appearance of a strong anisotropy in the MR of \Rb is consistent with the proposed quasi-one-dimensional character of band structure and its evolution with alkali metal species in this new family of superconductors.

\end{abstract}


\vskip 300 pt

\maketitle


The recent discovery of superconductivity in  A$_2$Cr$_3$As$_3$ (A =
K \cite{K-poly}, Rb \cite{Rb-poly} and Cs \cite{Cs-poly}) has
garnered significant attention due to the possible presence of
reduced dimensionality in both normal and superconducting state
properties. Superconductivity in quasi-one-dimensional (quasi-1D)
systems presents a unique set of properties, including exotic
deviations from expected behavior such as found in orbital magnetic
field properties, as well as unconventional pairing in the form of
singlet $d$-wave \cite{singlet} and triplet $p$-wave \cite{triplet}
order parameters that emerge.

\begin{figure}[!t]
\centering
\includegraphics[width=0.45 \textwidth]{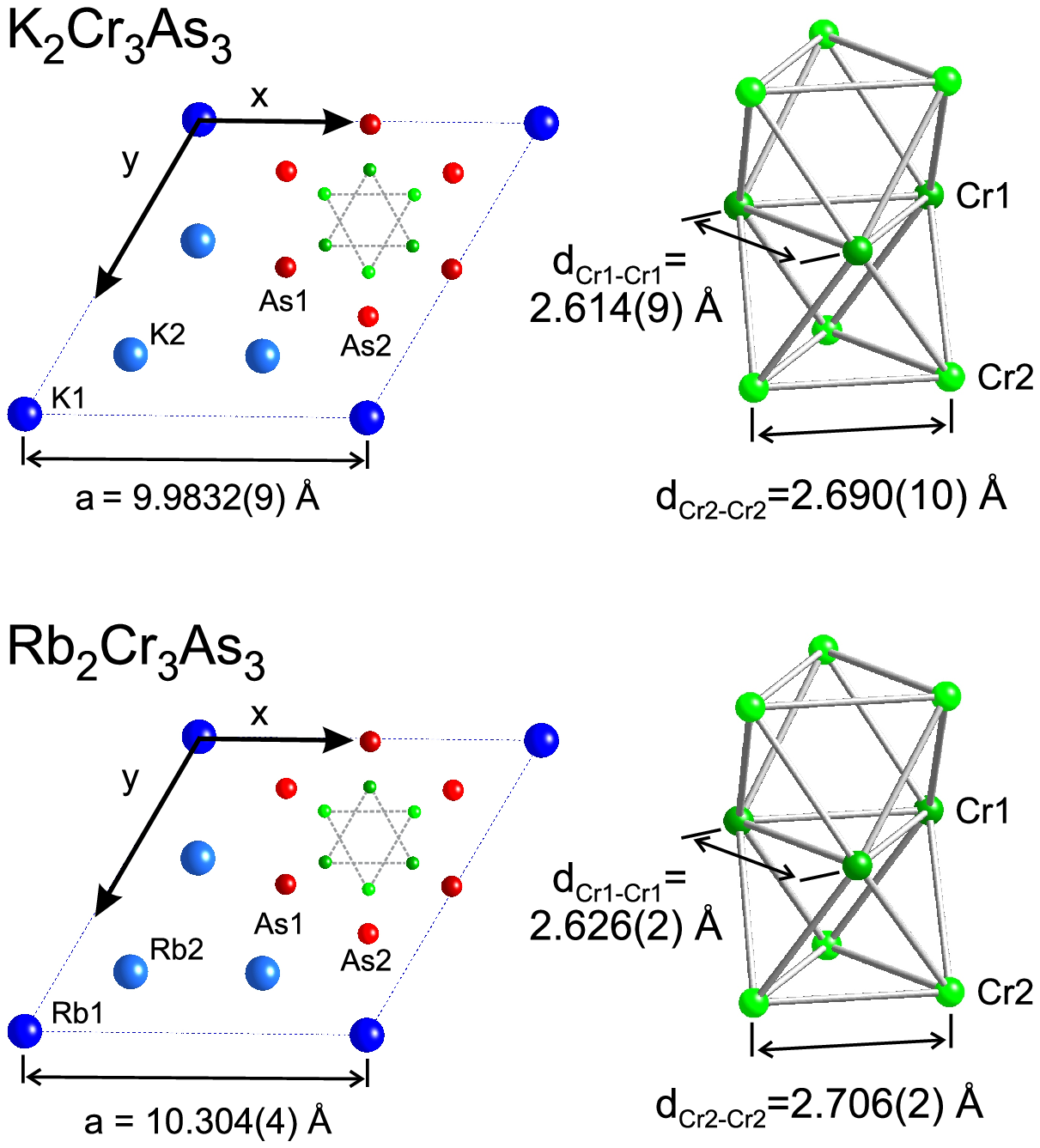}
\caption{Schematic view of structure of \K and \Rb. The detail
comparision of the lattice parameters are listed in Table \RNum{2}.
Data for \K is from Ref.~\onlinecite{K-poly}.}\label{Illustrate}
\end{figure}

The crystal structure of \A indeed has one-dimensional elements, composed of
double-walled subnanotube [Cr$_3$As$_3$]$_\infty$ chains separated
by alkali metal atoms in a hexagonal unit cell, as shown in Fig.~\ref{Illustrate}.
This chain crystal
structure is quite similar to the well known quasi-1D
superconductors such as the Bechgaard salts \cite{BechSalt,
BechSalt2}, and purple bronzes Li$_{0.9}$Mo$_6$O$_{17}$ \cite{LMO} and
Tl$_2$Mo$_6$Se$_6$ \cite{TMS}. The reported superconducting state
properties show some indications of reduced dimensions, including a
very large upper critical field \Hc that exceeds the Pauli limit
\cite{K-poly, K-single, Rb-poly}, and provide some evidence for an
unconventional pairing mechanism, as evidenced by power law behavior
in specific heat ~\cite{Rb-poly} and penetration depth measurements
\cite{Penetration} on polycrystalline samples, as well as strong
magnetic fluctuations in nuclear magnetic resonance \cite{K-NMR}
measurements.
However, due to the difficulties in interpreting measured properties of
polycrystalline samples with potential low-dimensional aspects, it
is imperative that single-crystalline samples are studied. For
instance, reports of an unusual linear temperature dependence of
resistivity in polycrystalline samples of \K and \Rb
\cite{K-poly,Rb-poly} have been contradicted by a recent
single-crystal study of \K \cite{K-single}, which also reports a
relatively weak anisotropy of the amplitude of \Hc between fields applied parallel ($H_{\parallel}$) and
perpendicular ($H_{\perp}$) to the needle-like crystal
orientation,which is the (00\emph{l}) direction of the crystal
structure, raising the question about the dimensionality of the
electronic structure in the A$_2$Cr$_3$As$_3$ family.

Here we report the successful growth of both \K and \Rb single
crystals and their normal and superconducting state properties based
on structural, transport, magnetic and thermodynamic measurements.
Comparing normal state transport properties between \K and \Rb
single crystal samples reveals several key differences, including a
large increase in the electrical conductivity in \Rb consistent with
the proposed band structure evolution, as well as the emergence of a
strong magnetoresistance magnitude and anisotropy in \Rb that we relate to a tunable reduced structural dimensionality that is controlled by the alkali metal species in the \A system.

\K and \Rb single crystals were grown using a self-flux method as
previously reported \cite{K-poly, K-single}. Pieces of K and Rb alkali metal
(99.8\%, Alfa Aesar) and chromium (99.999\%, Alfa Aesar) were
combined with arsenic powder (99.99\%, Alfa Aesar) into an alumina
crucible and sealed in a nitrogen glove box using a stainless steel
tube/swagelok cap configuration \cite{KFe2As2}. Growths were heated
up to 1000~$^o$C and kept at this temperature for one day, followed
by slow cooling down to 650~$^o$C. The swagelok enclosure was then
re-opened in a glovebox, revealing many single crystals with shiny,
silver coloring and a thin, long needle-like shape that extends
along the crystallographic [001] axis. Crystals were mechanically removed from the crucibles and visually inspected for residual flux and crystalline quality. We note that \Rb crystals are much more reactive than \K, resulting in larger uncertainties in sample mass determinations. 
Resistivity measurements were
performed using the standard four-terminal configuration in a
commercial cryostat system. Contacts were made in a glovebox using
silver paint, followed by coating of samples with Apiezon N-grease
to avoid air exposure. Magnetization measurement were taken using
both a commercial SQUID magnetometer and vibrating sample
magnetometer for \K and \Rb samples, respectively. A needle-like
specimen of \Rb with approximate dimensions 0.01~mm $\times$~0.03~mm
$\times$~0.39~mm was used for x-ray crystallographic analysis with
intensity data measured using a Bruker APEX-II CCD system equipped
with a graphite monochromator and a MoKa sealed tube ($\lambda$ =
0.71073~$\AA$), with data collection performed at 90~K and 250~K.
The structure was solved and refined using the Bruker SHELXTL
Software Package, using the space group P$\overline{6}$m2, with Z =
2 for the formula unit, \Rb. The final anisotropic full-matrix
least-squares refinement on F2 with 23 variables converged to $R1$
(w$_{R2}$) = 3.39\% (8.37\%) and 3.85\% (9.96\%) at 90~K and 250~K,
respectively.


\begin{figure}[!t]
\centering
\includegraphics[width=0.5 \textwidth]{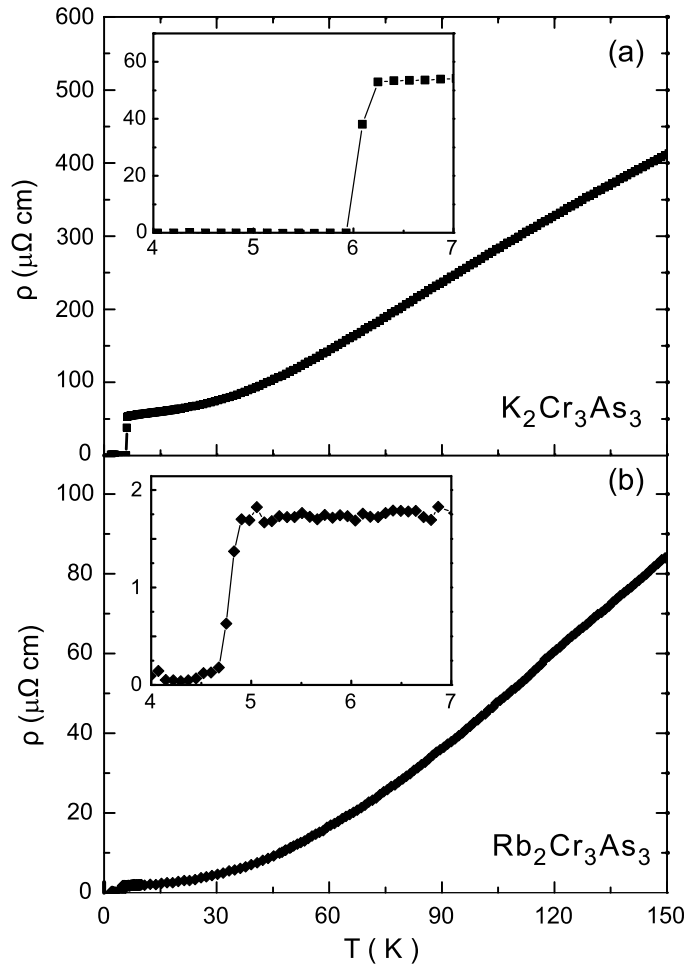}
\caption{Resistivity of \K and \Rb single crystals measured along the
[001] crystallographic (needle) direction, shown in panels (a) and (b), respectively.
Insets present low-temperature zoom, highlighting the superconducting transition in each compound.} \label{rho}
\end{figure}

\begin{figure}[!t]
\centering
\includegraphics[width=0.45 \textwidth]{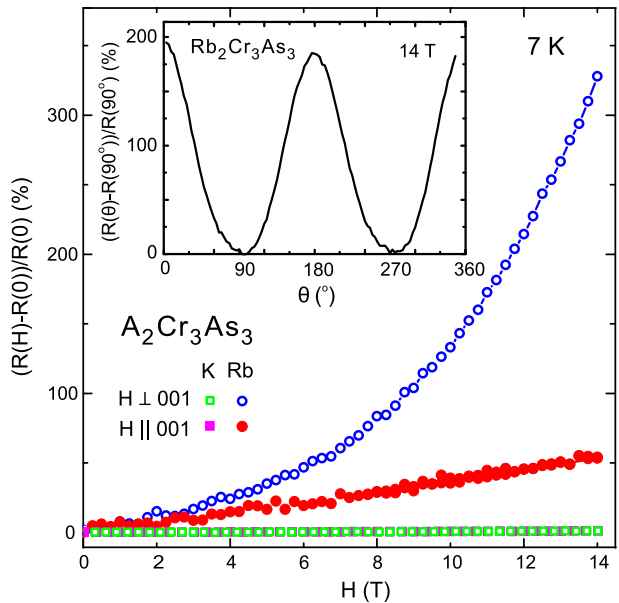}
\caption{Normalized magnetoresistance of \K and \Rb single crystal samples measured at 7~K under different field orientation configurations, with current applied along the needle ([001]) direction. The inset presents the field angle-dependent
resistivity of \Rb at 7~K and 14~T.} \label{MR}
\end{figure}

\begin{table}
\caption{Atomic coordinates for \Rb single crystal, measured at 250~K.} \label{tab:abc}
\begin{tabular}{lllllcc}
\hline
\bf Atom &\bf Label &\bf x&\bf y&\bf z&\bf Wyckoff&\bf U$_{eq}$ \\
&&&&&\bf position&\\
 \hline
\bf Rb&Rb1  &0.0&   0.0&    0.0&    3j&0.0279(10)\\
\bf Rb&Rb2& 0.79519(14)&    0.5904(3)&  0.5&3k  &0.0232(5)\\
\bf Cr&Cr1  &0.4183(2)& 0.8366(4)&  0.5&    3k&0.0115(7)\\
\bf Cr&Cr2& 0.5084(5)&  0.7542(2)&  0.0 &3j&0.0122(7)\\
\bf As&As1& 0.49505(14)&    0.9901(3)&  0.0&1c& 0.0125(5)\\
\bf As&As2  &0.6581(3)& 0.82907(15)&    0.5&3k& 0.0124(5)\\
 \hline
\end{tabular}
\end{table}

The lattice parameters of \Rb are larger than those of \K, with the
larger ionic radius of Rb (1.52~\AA) as compared to K (1.38~\AA)
increasing the spacing between the [Cr$_3$As$_3$]$_\infty$ chains as
shown in Fig.~\ref{Illustrate}. Table~\ref{tab:abc} presents the
atomic coordinates obtained from single-crystal x-ray analysis of
\Rb, with results well refined using a hexagonal crystal structure
(\SP) and showing no symmetry changes when cooling down to 90~K. The
unit cell of \Rb is 3.2\% and 0.5\% larger than that of \K
\cite{K-poly} along the $a$- and $c$-axis directions, respectively.
However, in contrast to the chain structure distortion previously
reported for polycrystalline samples \cite{K-poly}, the
[(Cr$_{3}$As$_{3})^{2-}]_{\infty}$ chain trunk width is almost
identical for \Rb and \K as shown in Table~\ref{table2}, displayed
more clearly in Fig.~\ref{Illustrate}. The Cr1-Cr1, Cr1-Cr2 and
Cr2-Cr2 bond distances change less than 0.6\% between \K and \Rb,
while the alkali metal-arsenic bonds change by 3-4\%, consistent
with a more simple expansion of the interchain distances with larger
alkali metal units and almost no influence on the chain structure
itself. This presents a unique ability to tune  the electronic
density of a system while keeping the important structural elements
unchanged, providing for an interesting ``knob'' to use for studying
the emergence of reduced dimensional properties.

\begin{table}
\caption{Comparision of crystallographic data for \K (300~K) \cite{K-poly} and
\Rb (250~K and 90~K).} \label{tab:xyz}
\begin{tabular}{llll}
\hline  &\bf \K\cite{K-poly}&\bf \Rb&\bf \Rb\\
 &300~K&250~K&90~K\\
\hline
 \bf space group&\bf P$\overline{6}$m2&\bf P$\overline{6}$m2&\bf P$\overline{6}$m2\\
\bf a($\AA$)&9.9832(9)&10.304(4)&10.251(5)\\
\bf c($\AA$)&4.2304(4)&4.2514(18)&4.227(2)\\
\bf V($\AA$$^3$)&365.13(6)&390.9(4)&384.7(4)\\
 \hline
\bf Bond distance&(A=K and Rb )&&\\
\bf A1-As1($\AA$) &3.366(2)&3.484(2)&3.462(2)\\
\bf A1-As2($\AA$)&3.263(5)&3.393(2)&3.375(2)\\
\bf A2-As1($\AA$)&4.9916(13)&5.153(2)&5.126(2)\\
\bf A2-As2($\AA$)&3.562(3)&3.718(2)&3.693(2)\\
\bf Cr1-As1($\AA$)&2.516(4)&2.529(2)&2.516(2)\\
\bf Cr1-As2($\AA$)&2.510(3)&2.511(2)&2.504(2)\\
\bf Cr2-As1($\AA$)&2.490(3)&2.502(2)&2.494(2)\\
\bf Cr2-As2($\AA$)&2.506(4)&2.511(2)&2.497(2)\\
\bf Cr1-Cr1($\AA$)&2.614(9)&2.626(2)&2.616(2)\\
\bf Cr1-Cr2($\AA$)&2.6116(15)&2.625(2)&2.613(2)\\
\bf Cr2-Cr2($\AA$)&2.690(10)&2.706(2)&2.703(2)\\
 \hline
\label{table2}
\end{tabular}
\end{table}

The electrical resistivity of \K and \Rb single crystals along the
[001] needle direction is presented in Fig.~\ref{rho}, showing \Tc
values of 6.1~K and 4.8~K, respectively, consistent with previous
reports \cite{K-poly,K-single,Rb-poly}.
Transport behavior of \K crystals is quite similar to that
previously reported for single-crystal samples \cite{K-single},
showing a marked difference in the temperature dependence as
compared to polycrystalline samples that exhibit an anomalous linear
temperature dependence above \Tc \cite{K-poly,Rb-poly,Cs-poly}.
The residual resistivity ratio $RRR$=$R$(300~K)/$R$(7~K) for \K is
15, slightly larger than the reported value of 10 for polycrystals
\cite{K-poly,K-single} and smaller than in previous work
\cite{K-single}. For the latter study, the reported ratio is 50
while the residual resistivity of $\rho_0 \simeq 25~\mu\Omega$cm is
about half of the value measured in this study, indicating a notable
sample dependence of resistivity without a notable change in \Tc.
For single-crystal \Rb synthesized in the exact same manner as \K,
we obtain a much larger RRR value of 125 as compared to \K
\cite{K-poly,K-single} and \Rb \cite{Rb-poly} studies, likely
indicating very high crystalline purity in this system and a
possible sensitivity to grain boundary effects as encounted in
polycrystalline studies.

Interestingly, the resistivity ($\sim$200 $\mu\Omega$cm) of
single-crystal \Rb at 300~K is notably reduced compared to that of
\K, by a factor of approximately four. Assuming the phonon scattering is similar
in the two compounds, this decrease may be a result of qualitative
changes in the electronic structure, consistent with the theoretical
prediction of significant changes in the Fermi surface by J.~Wu
\emph{et al} \cite{233-theory}, who point out that an extra quasi-1D
surface near [0 0 $\pi$] and changes in the three-dimensional
surfaces could result in an increase in conductivity in \Rb as
compared to \K.
We also observe striking contrasts in the magnetoresistance (MR) of
\K and \Rb crystals. The extremely small MR measured in \K ($<$1\%
at 7~K and 14~T) is similar to that found in previous work
\cite{K-single}, with no measurable difference between the MR
measured in both transverse ($H\bot [001]$) and longitudinal
$H\parallel [001]$) configurations up to 14~T. This is in stark
contrast to the MR of \Rb (Fig.~\ref{MR}), which exhibits a 325\%
increase in $H\bot [001]$ MR at 7~K and 14~T that is quasi-parabolic
and linear for \Hperp and \Hpar orientations, respectively.
Moreover, the anisotropy exhibits a two-fold symmetry in field-angle
rotation measurements shown in Fig.~\ref{MR}, with a 200\% change in
magnitude between \Hperp and \Hpar orientations. Considering the
similarity of the two systems in all other known properties, this
dramatic increase in MR and appearance of strong anisotropy is
considered strong evidence for the appearance of quasi-1D components
in the electronic structure, consistent with theoretical predictions
as noted above \cite{233-theory}.

\begin{figure}[!t]
\centering
\includegraphics[width=0.45 \textwidth]{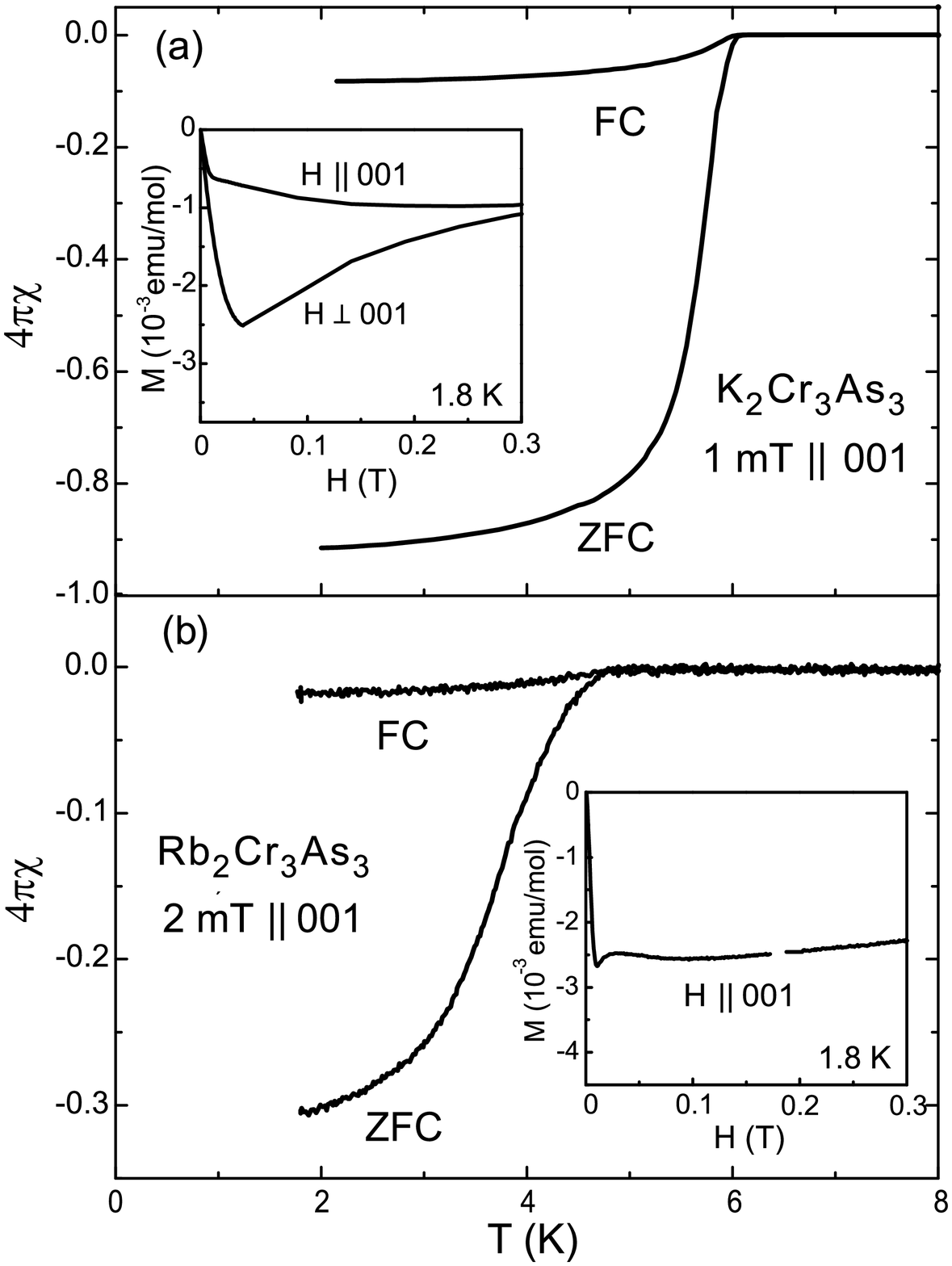}
\caption{Magnetic susceptibility of \K and \Rb single crystal collections under zero-field cooled and field-cooled conditions.
Panel (a) presents measurements of \K single
crystals with 1~mT field applied along the [001] needle direction.
The insert shows field dependent magnetization of \K single crystals
with field applied both parallel and perpendicular to [001].
Panel (c) presents measurements of \Rb single
crystals with 2~mT field applied along the [001] needle direction. The insert shows the field dependent magnetization for parallel (\Hpar) fields.} \label{magnetization}
\end{figure}

\begin{figure}[!t]
\centering
\includegraphics[width=0.5 \textwidth]{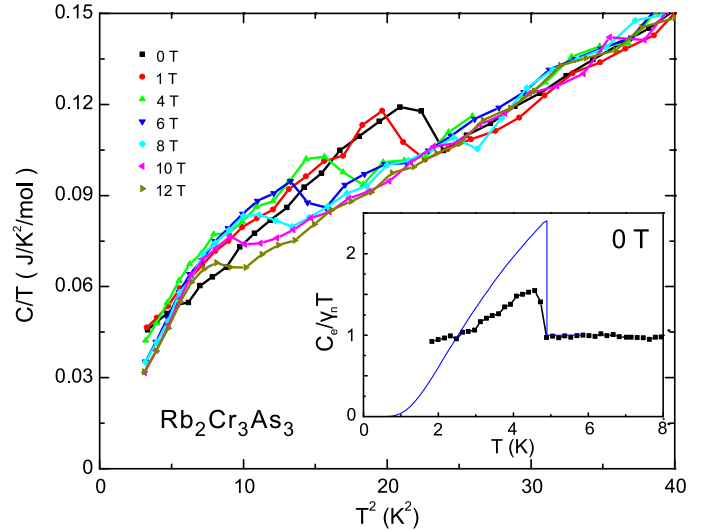}
\caption{Heat capacity of \Rb single crystals with H$\bot$[001]. The
inset shows the normalized electronic specific heat
C$_{e}$/($\gamma$$_{n} T$) as a function of temperature; blue line is the theoretical BCS expectation for a single-band $s$-wave superconductor (see text). } \label{spheat}
\end{figure}

The magnetic susceptibility $\chi$ of \K and \Rb single crystals,
shown in Fig.~\ref{magnetization}, confirms the onset of diamagnetic
Meissner screening at the same \Tc values reported above. In \K, a
shielding fraction of $\sim$90\% is estimated using the
zero-field-cooled $\chi(T)$ curve, indicating a significant bulk
fraction of superconductivity. In \Rb, a smaller apparent fraction
of $\sim$30\% may be due to difficulties in measuring the much
smaller crystals, with needle diameters of 30~$\mu$m as compared to
80~$\mu$m for \K crystals, and/or due to greater errors in mass determination because of the much more reactive nature of the compound.
The isothermal magnetization $M$ was
measured for both systems at 1.8~K, as shown in the insets of
Fig.~\ref{magnetization}. The low-field behavior exhibis the
characteristic curve of type \RNum{2} superconductors, where
$H_{c1}$ can be obtained from the deviation point of the linear
field dependence of the $M(H)$ curve, yielding ($H\parallel$[001])
values of 90~Oe and 60~Oe in \K and \Rb, respectively.

\begin{figure}[!t]
\centering
\includegraphics[width=0.5 \textwidth]{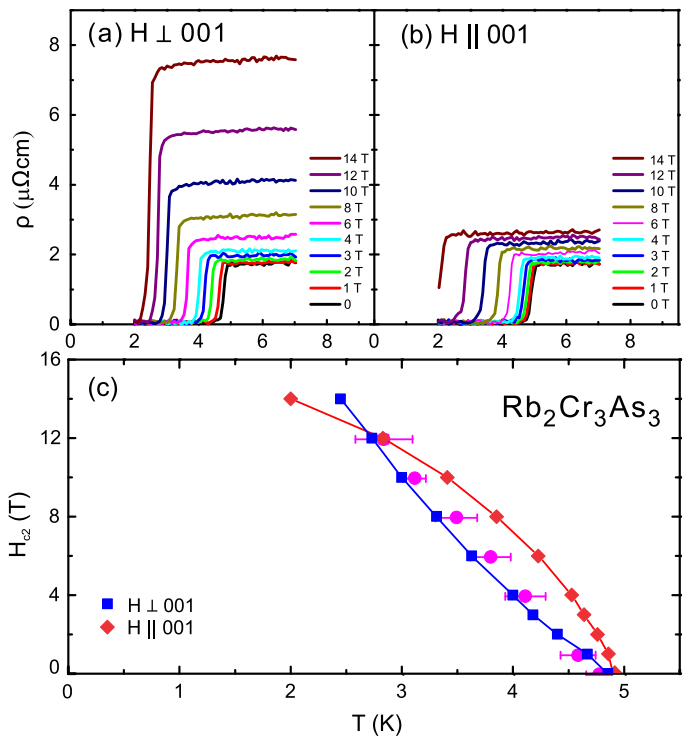}
\caption{Upper critical field of \Rb single crystal under perpendicular (panel (a)) and parallel (panel (b)) field orientations. Temperature dependence of \Hc from resistance data is shown in panel (c), with $T_c(H)$ defined by the midpoint between zero and full resistance. This is compared with \Tc(H) data obtained from heat capacity with $H\bot [001]$ orientation (solid purple circles), with \Tc values defined by the midpoint of the jump in specific heat and error bars corresponding to the transition width.}
\label{Hc2-Rb}
\end{figure}

Bulk superconductivity in \Rb is also confirmed by heat capacity
measurements on a collection of single crystals (total mass $\sim$
0.2~mg) oriented with $H\bot [001]$, showing a clear jump in
specific heat $C$ at \Tc as shown in Fig.~\ref{spheat}. The jump is
found to decrease in temperature as magnetic field is increased,
indicating the suppression of a bulk superconducting phase
transition with field. \Tc obtaining from the middle point of the
jump were plotted in Fig.~\ref{Hc2-Rb}, which shows remarkably
consistency with the transport data. Fitting the normal state heat
capacity with the expression $C/T$=$\gamma$$_n$+$\beta$T$^2$, yields
values for the electronic Sommerfeld coefficient $\gamma$$_n$ =
39.5~mJ K$^{-2}$ mol$^{-1}$ and the phonon coefficient $\beta$ =
2.8~mJ K$^{-4}$ mol$^{-1}$. The value for $\gamma$$_n$ is somewhat
smaller than that reported for polycrystalline samples (55.1~mJ
K$^{-2}$ mol$^{-1}$) \cite{Rb-poly}, while that of $\beta$ is close
to previous reports.
We extract the electronic component $C_e$ using these parameters and
plot its temperature dependence as C$_e$/($\gamma$$_n$T) in the
inset of Fig.~\ref{spheat}, showing the dimesionless heat capacity
jump at \Tc to be $\Delta C/\gamma_n T_c$=0.62, much lower than the
reported values of 1.8 in \Rb \cite{Rb-poly} and 2.0-2.4 in \K
\cite{K-poly, K-single}). It is also lower than the BCS value of
1.43 expected for a conventional superconductor, which could be
related to the apparent reduced superconducting volume fraction
observed in susceptibility measurements noted above, or more
intrinsic origins to do with the nature of superconductivity that
require further investigation at lower temperatures.

\begin{figure}[!t]
\centering
\includegraphics[width=0.5 \textwidth]{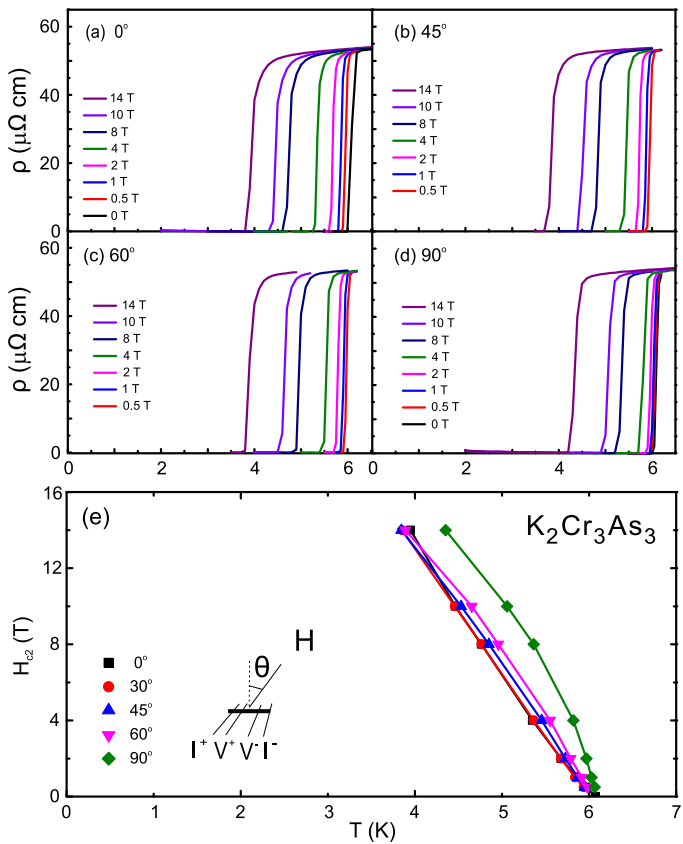}
\caption{Upper critical field of \K single crystal under different field
configurations. Panels (a), (b), (c) and (d) present the field evolution of temperature-dependent resistivity under different field orientations with $\theta$ = 0$^o$, 45$^o$,
60$^o$ and 90$^o$, respectively.  Panel (e) shows the upper critical
field temperature dependence in different field orientations as extracted from resistivity transitions in panels (a)-(d), with $T_c(H)$ defined by the midpoint between zero
and full resistance.} \label{Hc2-K}
\end{figure}

\begin{figure}[!t]
\centering
\includegraphics[width=0.45 \textwidth]{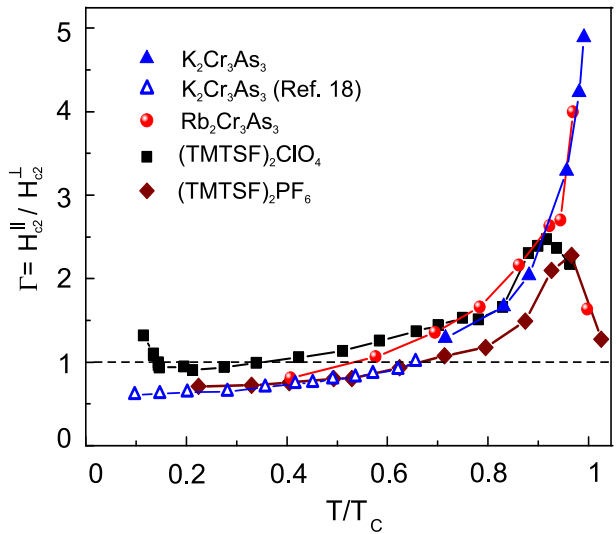}
\caption{Anisotropy of the upper critical field $\Gamma \equiv H_{c2}^{\parallel}$(H$\parallel$00\emph{l} for 233,
H$\parallel$a for TMTSF salts)/$H_{c2}^{\bot}$(H$\perp$00\emph{l} for 233,
H$\parallel$b for TMTSF salts). \Tc of \K and \Rb are defined using
the midpoint between zero and full resistance. The data for TMTSF
salt is taken from Refs.~\cite{TMTSFClO6} and \cite{TMTSFPF6}, and high-field data for \K is from Ref.~\cite{K233-highfield}.}
\label{Hc2norm}
\end{figure}

In order to understand the extent of influence of quasi-1D band
structure components on the superconducting state, the anisotropy of
\Hc was investigated in both \K and \Rb. The resistive transition
was tracked as a function of field angle, as shown in
Figs.~\ref{Hc2-K} and \ref{Hc2-Rb}, using orientation angles
$\theta$ = 0$\degree$, 45$\degree$, 60$\degree$ and 90$\degree$ from
the \Hperp orientation for \K and $\theta$ = 0$\degree$ and
90$\degree$ for \Rb. For \K, the initial slope $dH_{c2}(T_c)/dT$
varies from 5.0~T/K to 16.1~T/K, as observed previously
\cite{K-single}, with a rapid increase that is even more significant
than what were observed in other quasi-1D superconducting state, as
shown in Fig.~\ref{Hc2norm}.
For \Rb, the initial slope $dH_{c2}(T_c)/dT$ has a similar
range between \Hperp and \Hpar orientations, with values of 4.4~T/K and 14~T/K, respectively. 
However, the temperature dependence of $H_{c2}(T)$ in \Rb is quite different
than that of \K in the field range up to 14~T, showing a positive
curvature for \Hperp and negative curvature for \Hpar as shown in
Fig.~\ref{Hc2-Rb}.
In fact, the $H_{c2}(T)$ curves for the two field orientations cross
near $\sim$12~T, in a manner similar to that observed in other
classic quasi-1D superconductors such as \ClO \cite{TMTSFClO6} and
\PF \cite{TMTSFPF6}. In these organic superconductors, a similar
trend is observed where a diverging temperature dependence is
observed in the \Hperp orientation, while a saturating Pauli-limited
dependence occurs for \Hpar (parallel to the conducting direction). In
Li-based purple bronze, such Pauli-limited behavior can be also observed
with fields applied along the conducting chain, while a divergent behavior is observed in the other two field orientations \cite{HC2LMO}.

Recent pulsed-field work on \K single crystals indeed shows such a crossing to occur at higher fields, with $H_{c2}^{\bot}(T)$ surpassing $H_{c2}^{\parallel}(T)$ near 15~T and indications of Pauli-limited behavior in the \Hpar orientation \cite{K233-highfield}.
Fig.~\ref{Hc2norm} presents a comparison between the \A compounds
and the organic systems \ClO and \PF, plotting the temperature
dependence of the \Hc anisotropy parameter
$\Gamma(T)=H_{c2}^{\parallel}(T)/H_{c2}^{\bot}(T)$ (with \Hpar and \Hperp
relative to the $c$-axis for \A, and in the directions of $a$- and
$c$-axis directions, respectively, for TMTSF salts). A similar
change in $\Gamma$, from large near \Tc to near unity at lower
temperatures, is seen in both the organics and \A materials,
indicating a resemblance. Although a similar $\Gamma(T)$ evolution
was also found in higher-dimensional systems such as K-doped
BaFe$_2$As$_2$ \cite{Altarawneh,Yuanhq}, the notable difference is in the
crossing (\ie, $\Gamma$=1) point, where the \Hperp component begins
to take off in the quasi-1D systems. While a mild anisotropy was
previously reported for \K single crystals \cite{K-single}, it is
clear that the observable trend in the limited field range available
for our measurements of \K is very similar to that of \Rb, and indeed follows the same behavior at higher fields \cite{K233-highfield} as shown in Fig.~\ref{Hc2norm} for a normalized temperature regime.

Although higher field data for \Rb is lacking, the apparent likeness of $\Gamma(T)$ between \Rb and \K presents an apparent contradiction: contrasting MR anisotropies are suggestive of a highly tunable quasi-1D normal state electronic structure that differs in the two compounds, while the normalized $\Gamma(T)$ behavior are very comparable for each. This conundrum could be readily dismissed if Pauli limiting were the dominant pairbreaking mechanism, as the comparison of superconducting condensation energy and Zeeman energy is, to first order, independent of details of the electronic band structure of the host material. However, the observed combination of orbital (\Hperp) and Pauli (\Hpar) limited \Hc behaviors in \K \cite{K233-highfield} calls for a less simplified explanation. For instance, the spin-locked Cooper pairing scenario proposed by Balakirev {\it et al.}, where the pair singlet spins are pinned along the quasi-1D chain direction \cite{K233-highfield}, may provide a pairbreaking scenario common to all \A superconductors, given their pairing attraction is indeed tied to a low-dimensional spin fluctuation mechanism in all three systems. Above all, this highlights the need for detailed experimental studies of the electronic structure as well as higher-field \Hc studies of \Rb in order to understand how the anisotropy of superconducting state properties is affected by the strongly enhanced normal state anisotropy found in \Rb.

In summary, we report successful synthesis of high-quality \A single
crystals with A = K and Rb using a self-flux method, providing the
first comparison of single-crystal properties between two members of
this new superconductor family. The increase in ionic size from K to
Rb causes an increase in the spacing between quasi-one-dimensional
chains in the hexagonal crystal structure of \Rb as compared to \K,
while keeping the intra-chain dimensions mostly unchanged. While the
superconducting state properties are comparable between the two
systems, the normal state resistivity shows a striking evolution,
with a 325\% enhancement of magnetoresistance and appearance of
strong anisotropy in \Rb. These properties point to a strong quasi-one-dimensional component in the electronic properties that evolves with enhanced unit cell dimensions in the \A system. The similarities to other well-established quasi-1D superconductors
deserves further attention.

The authors acknowledge
R. L. Greene,
H. Hodovanets,
J. P. Hu,
Y. P. Jiang,
L. M. Wang,
and X. H. Zhang
for valuable discussion and experimental assistance.
This work was supported by AFOSR through Grant FA9550-14-1-0332 and
the Gordon and Betty Moore Foundation's EPiQS Initiative
through Grant GBMF4419.

\end{document}